\documentclass[
    ,final            
  ]
  {aipproc}

\layoutstyle{8x11single}

\begin{document}

\title{Observationally Verifiable Predictions of Modified Gravity\footnote{Talk given by JWM at the ``The Invisible Universe'' conference, Paris, France, June 29--July 3, 2009.}}


\author{J.~W. Moffat}{
  address={Perimeter Institute for Theoretical Physics, Waterloo, Ontario N2L 2Y5, Canada}
  ,altaddress={Department of Physics, University of Waterloo, Waterloo, Ontario N2L 3G1, Canada}
}

\author{V.~T. Toth}{
  address={Perimeter Institute for Theoretical Physics, Waterloo, Ontario N2L 2Y5, Canada}
}

\begin{abstract}
MOG is a fully relativistic modified theory of gravity based on an action principle. The MOG field equations are exactly solvable numerically in two important cases. In the spherically symmetric, static case of a gravitating mass, the equations also admit an approximate solution that closely resembles the Reissner-Nordstr\"om metric. Furthermore, for weak gravitational fields, a Yukawa-type modification to the Newtonian acceleration law can be obtained, which can be used to model a range of astronomical observations. Without nonbaryonic dark matter, MOG provides good agreement with the data for galaxy rotation curves, galaxy cluster masses, and gravitational lensing, while predicting no appreciable deviation from Einstein's predictions on the scale of the solar system. Another solution of the field equations is obtained for the case of a a spatially homogeneous, isotropic cosmology. MOG predicts an accelerating universe without introducing Einstein's cosmological constant; it also predicts a CMB acoustic power spectrum and a mass power spectrum that are consistent with observations without relying on non-baryonic dark matter. Increased sensitivity in future observations or space-based experiments may be sufficient to distinguish MOG from other theories, notably the $\Lambda$CDM ``standard model'' of cosmology.
\end{abstract}

\maketitle


\section{Introduction}

MOG is a fully relativistic modified theory of gravity, also known as Scalar-Tensor-Vector Gravity (STVG \cite{Moffat2006a,Moffat2007e}. The theory is derived from an action principle. It is a stable, self-consistent gravity theory that can describe solar system, astrophysical, and cosmological data.

The principal feature of the STVG theory is the extra degrees of freedom introduced through a massive vector field called a ``phion'' field. The curl of the phion field is a skew-symmetric tensor field that couples to matter, realizing a repulsive ``fifth force'' interaction. This fifth force acts in combination with the gravitational field (described by the usual symmetric Einstein metric tensor) and obeys the weak equivalence principle. At short range, the presence of the repulsive force results in the observed value of Newton's gravitational constant, $G_N$. At intermediate ranges, the repulsive force begins to vanish, resulting in the flattened rotation curves of galaxies. The repulsive force vanishes completely at large distances, leading to a Newtonian type gravitational attraction with a gravitational constant larger than $G_N$.

In addition to the vector field, MOG allows for the gravitational constant $G$, the mass of the vector field $\mu$, and the vector field coupling constant $\omega$ to be running. In the weak field, non-relativistic approximation this amounts to the running of the effective gravitational constant and the range of the fifth force as functions of the source mass.

MOG should not be confused with Milgrom's Modified Newtonian Dynamics (MOND), or its relativistic generalization by Bekenstein, called Tensor-Vector-Scalar gravity (TeVeS). Apart from superficial similarities, the two theories have very little resemblance and offer different predictions.

In this paper, we review the foundations of MOG and show how the modified Newtonian acceleration law for weak fields is derived. We demonstrate that this acceleration law can fit, {\bf without nonbaryonic dark matter}, a large amount of galaxy rotation curve data, galaxy cluster data, and also explain gravitational lensing by the Bullet Cluster. We show that the theory can also explain the acoustic power spectrum of the cosmic microwave background, is consistent with Type Ia supernova data, and provides a model for structure formation that is consistent with the matter power spectrum. These results offer verifiable predictions that may be used in the near future to distinguish MOG from competing theories, including $\Lambda$CDM.

\section{Modified Gravity (MOG)}

The STVG action \cite{Moffat2006a} is constructed by starting with the Einstein-Hilbert Lagrangian density that describes the geometry of spacetime:
\begin{equation}
{\cal L}_G=-\frac{1}{16\pi G}\left(R+2\Lambda\right)\sqrt{-g},
\end{equation}
where $G$ is the gravitational constant, $g$ is the determinant of the metric tensor $g_{\mu\nu}$ (we are using the metric signature $(+,-,-,-)$), and $\Lambda$ is the cosmological constant. We set the speed of light, $c=1$. The Ricci-tensor is defined as
\begin{equation}
R_{\mu\nu}=\partial_\alpha\Gamma^\alpha_{\mu\nu}-\partial_\nu\Gamma^\alpha_{\mu\alpha}+\Gamma^\alpha_{\mu\nu}\Gamma^\beta_{\alpha\beta}-\Gamma^\alpha_{\mu\beta}\Gamma^\beta_{\alpha\nu},
\end{equation}
where $\Gamma^\alpha_{\mu\nu}$ is the Christoffel-symbol, while $R=g^{\mu\nu}R_{\mu\nu}$.

We introduce a ``fifth force'' vector field $\phi_\mu$ via the Maxwell-Proca Lagrangian density:
\begin{equation}
{\cal L}_\phi=-\frac{1}{4\pi}\omega\left[\frac{1}{4}B^{\mu\nu}B_{\mu\nu}-\frac{1}{2}\mu^2\phi_\mu\phi^\mu+V_\phi(\phi)\right]\sqrt{-g},
\end{equation}
where $B_{\mu\nu}=\partial_\mu\phi_\nu-\partial_\nu\phi_\mu$, $\mu$ is the mass of the vector field, $\omega$ characterizes the strength of the coupling between the ``fifth force'' and matter, and $V_\phi$ is a self-interaction potential.

The three constants of the theory, $G$, $\mu$ and $\omega$, are promoted to scalar fields by introducing associated kinetic and potential terms in the Lagrangian density:
\begin{equation}
{\cal L}_S=-\frac{1}{G}\left[\frac{1}{2}g^{\mu\nu}\left(\frac{\nabla_\mu G\nabla_\nu G}{G^2}+\frac{\nabla_\mu\mu\nabla_\nu\mu}{\mu^2}-\nabla_\mu\omega\nabla_\nu\omega\right)+\frac{V_G(G)}{G^2}+\frac{V_\mu(\mu)}{\mu^2}+V_\omega(\omega)\right]\sqrt{-g},
\end{equation}
where $\nabla_\mu$ denotes covariant differentiation with respect to the metric $g_{\mu\nu}$, while $V_G$, $V_\mu$, and $V_\omega$ are the self-interaction potentials associated with the scalar fields.

The action integral takes the form
\begin{equation}
S=\int{({\cal L}_G+{\cal L}_\phi+{\cal L}_S+{\cal L}_M)}~d^4x,
\label{eq:FldL}
\end{equation}
where ${\cal L}_M$ is the ordinary matter Lagrangian density, such that the energy-momentum tensor of matter becomes
\begin{equation}
T_{\mu\nu}=-\frac{2}{\sqrt{-g}}\frac{\delta S_M}{\delta g^{\mu\nu}},
\end{equation}
where $S_M=\int{\cal L}_M~d^4x$. A ``fifth force'' matter current can be defined as:
\begin{equation}
J^\nu=-\frac{1}{\sqrt{-g}}\frac{\delta S_M}{\delta\phi_\nu}.
\end{equation}
We assume that the variation of the matter action with respect to the scalar fields vanishes:
\begin{equation}
\frac{\delta S_M}{\delta X}=0,
\end{equation}
where $X=G,\mu,\omega$.

From the test particle Lagrangian
\begin{equation}
{\cal L}_\mathrm{TP}=-m+\alpha\omega q_5\phi_\mu u^\mu,
\label{eq:TP}
\end{equation}
where $m$ is the test particle mass, $q_5$ is its fifth force charge which we find to be equal to $q_5=m\sqrt{G_N/\omega}$, $u^\mu=dx^\mu/ds$ is its four-velocity, and $\alpha$ is a factor representing the nonlinearity of the theory, we obtain the test particle equation of motion \cite{Moffat2006a,Moffat2007e}:
\begin{equation}
m\left(\frac{du^\mu}{ds}+\Gamma^\mu_{\alpha\beta}u^\alpha u^\beta\right)=-\alpha\kappa\omega mB^\mu{}_\nu u^\nu.
\end{equation}
The field equations derived from the STVG Lagrangian have exact numerical solutions \cite{Moffat2007e}. In the spherically symmetric, static case it is also possible to derive an approximate solution by writing the metric in the form
\begin{equation}
ds^2=Bdt^2-Adr^2-r^2d\Omega^2,
\end{equation}
in which case the solution takes the form
\begin{eqnarray}
\phi_t&=&-Q_5\frac{e^{-\mu r}}{r},\\
B&=&1-\frac{2(1+\alpha)G_NM}{r}+\frac{(1+\alpha)G_N^2M^2}{r^2},\\
A&=&B^{-1},\\
\mu&=&\frac{D}{\sqrt{M}},\\
\alpha&=&\frac{G_\infty-G_N}{G_N}\frac{M}{(\sqrt{M}+E)^2},
\end{eqnarray}
where $G_\infty\simeq 20G_N$ is determined by cosmological observations \cite{Moffat2007e,Moffat2007c}; $D\simeq 6250 M_\odot^{1/2}\mathrm{kpc}^{-1}$ and $E\simeq 25000 M_\odot^{1/2}$ are determined by fits to galaxy rotation curves; $M$ is the source mass at the center of the spherically symmetric, static field; and $Q_5=\kappa M$ is its fifth force charge. The constant $\kappa$ is given by $\kappa=\sqrt{G_N/\omega}$, where $\omega=1/\sqrt{12}$.

The fifth force charge $Q_5$ is proportional to the gravitating mass, thus guaranteeing the validity of the weak equivalence principle. This is possible because in Maxwell-Proca theory, charge is not conserved: $\nabla_\mu J^\mu\ne 0$.

\begin{figure}[t!]
\includegraphics[width=0.4\linewidth]{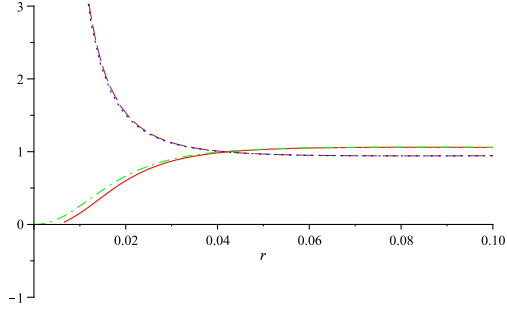}
\caption{Comparing MOG numerical solutions to the Reissner-Nordstr\"om solution, for a $10^{11}~M_\odot$ source mass. The MOG parameters $A$ (solid red line) and $B$ (dashed brown line) are plotted along with the Reissner-Nordstr\"om values of $A$ (dash-dot green line) and $B$ (dotted blue line). Horizontal axis is in pc. We observe that the $A$ parameter reaches 0 at below the Schwarzschild radius of a $10^{11}~M_\odot$ mass, which is $\sim 0.01$~pc. From \cite{Moffat2007e}.}
\label{fig:AB}
\end{figure}

This approximate solution is formally identical to the Reissner-Nordstr\"om solution of general relativity. Indeed, it can be shown by numerical analysis that this correspondence remains valid at extremely short radii (Fig.~\ref{fig:AB}). The MOG vacuum solution, interior matter solutions for stars and stellar collapse are currently being analyzed. Significant modifications of the standard black hole GR solution are expected to be discovered.

The effective gravitational field of a point source can be written in the form
\begin{equation}
\ddot{{\bf r}}=-\frac{G_\mathrm{eff}M{\bf r}}{r^3},~~~~~G_\mathrm{eff}=G_N\left[1+\alpha-\alpha(1+\mu r)e^{-\mu r}\right].
\end{equation}

We can also derive the MOG Poisson equation for the effective gravitational potential $\Phi$ in the form \cite{Moffat2007c,Brownstein2007}:
\begin{equation}
\nabla^2\Phi=4\pi G_N\rho({\bf r})+\alpha\mu^2 G_N\int\frac{e^{-\mu|{\bf r}-{\bf\tilde r}|}\rho({\bf\tilde r})}{|{\bf r}-{\bf\tilde r}|}d^3{\bf\tilde r}.
\end{equation}

\section{Fitting Galaxy Rotation Curves and Clusters}

\begin{figure}[t!]
\includegraphics[width=0.3\linewidth]{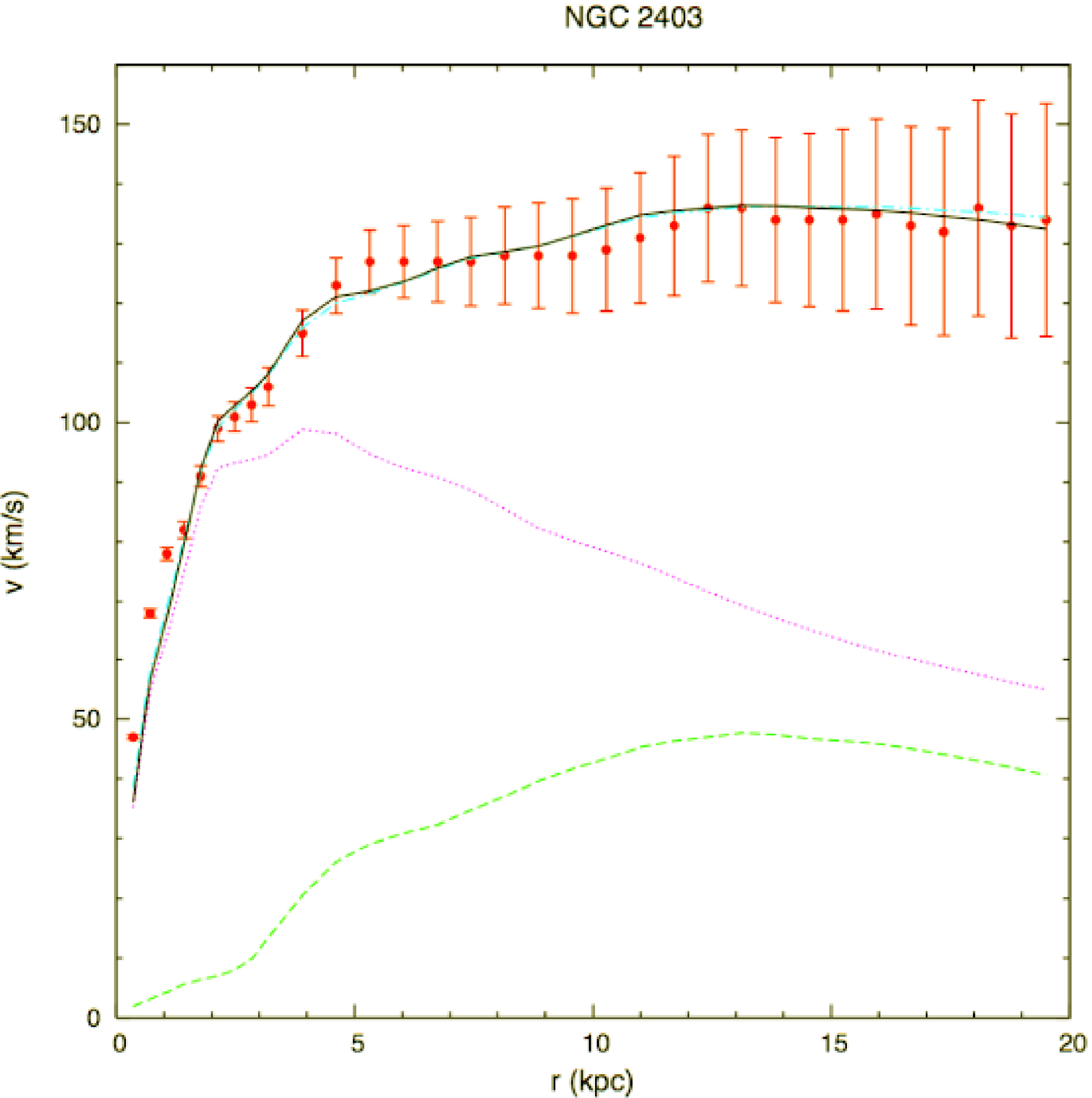}
\hskip 0.15\linewidth
\includegraphics[width=0.3\linewidth]{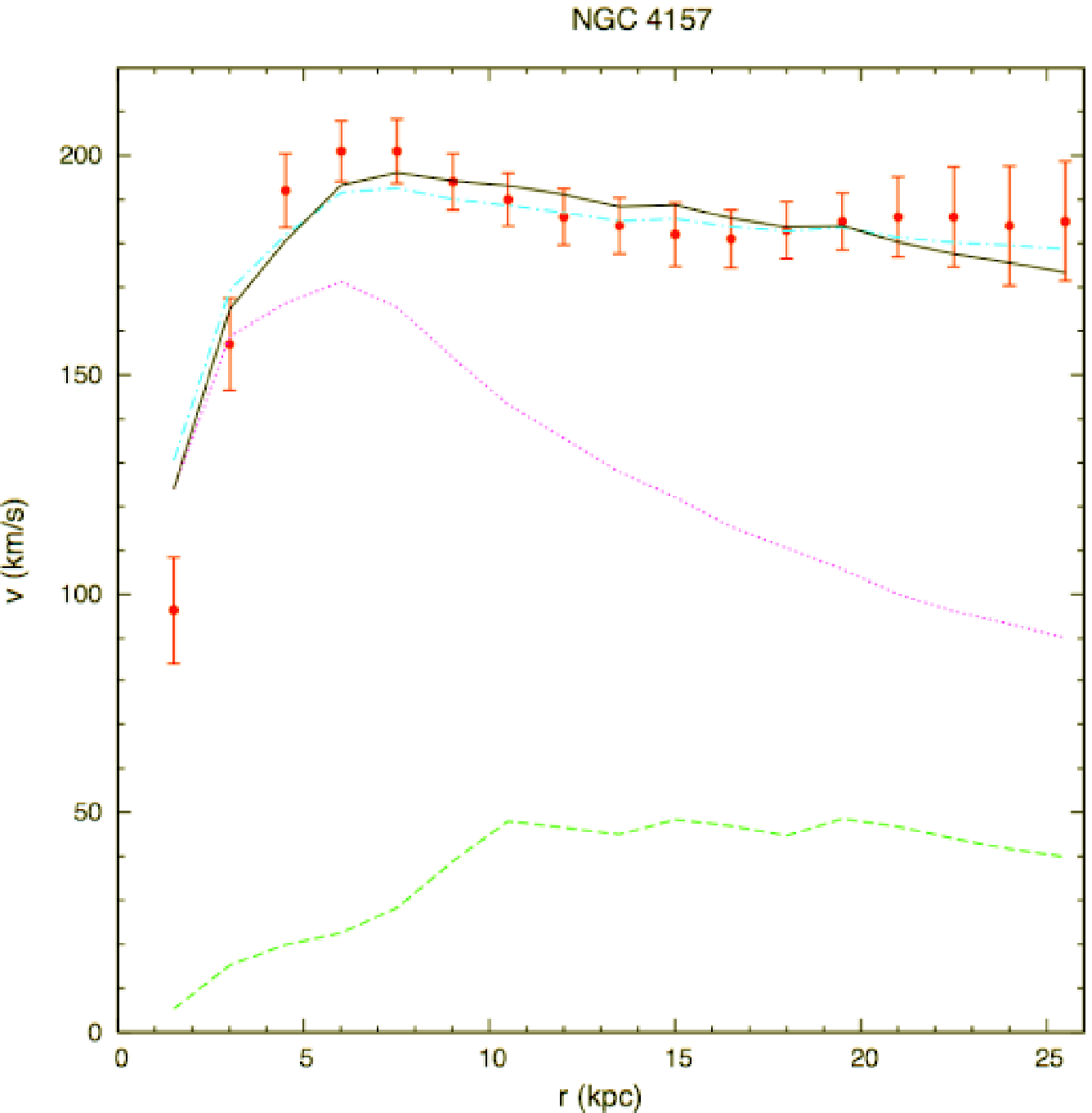}
\caption{Photometric fits to galaxy rotation curves. There are 2 benchmark galaxies presented here. Each is a best fit via the single parameter (M/L) based on the photometric data of the gaseous (HI plus He) and luminous stellar disks. The radial coordinate (horizontal axis) is given in kpc and the rotational velocity (vertical axis) in km/s. The red points with error bars are the observations, the solid black line is the rotation curve determined from MOG, and the dash-dotted cyan line is the rotation curve determined from MOND. The other curves are the Newtonian rotation curves of the various separate components: the long-dashed green line is the rotation curve of the gaseous disk (HI plus He) and the dotted magenta curve is that of the luminous stellar disk. From \cite{Brownstein:2009gz}.}
\label{fig:gals}
\end{figure}

Recently, we developed a fitting routine that was used to fit a large number of galaxy rotation curves (101 galaxies) using STVG (Fig.~\ref{fig:gals}). Of these, 58 galaxies were fitted using photometric data, while an additional 43 galaxies were fitted using a core model \cite{Brownstein:2009gz,Brownstein2006a}. {\bf For the photometric fits, only one parameter, the mass-to-light ratio (M/L) was used.} The fits are remarkably good for STVG, which is not surprising: the MOG solution can be used to derive the empirical Tully-Fisher law for galaxy rotation curves in the form \cite{Moffat2007e}:
\begin{equation}
v_r^2\propto\sqrt{M}.
\end{equation}
Furthermore, {\bf for every feature in the surface brightness distribution, MOG produces a corresponding feature in the predicted rotation curve (matching the observed rotation curve)}.

At large distances from a host galaxy (e.g., for satellite galaxies), the rotation curves again become Kepler-Newtonian.

\begin{figure}[t!]
\includegraphics[width=0.24\linewidth]{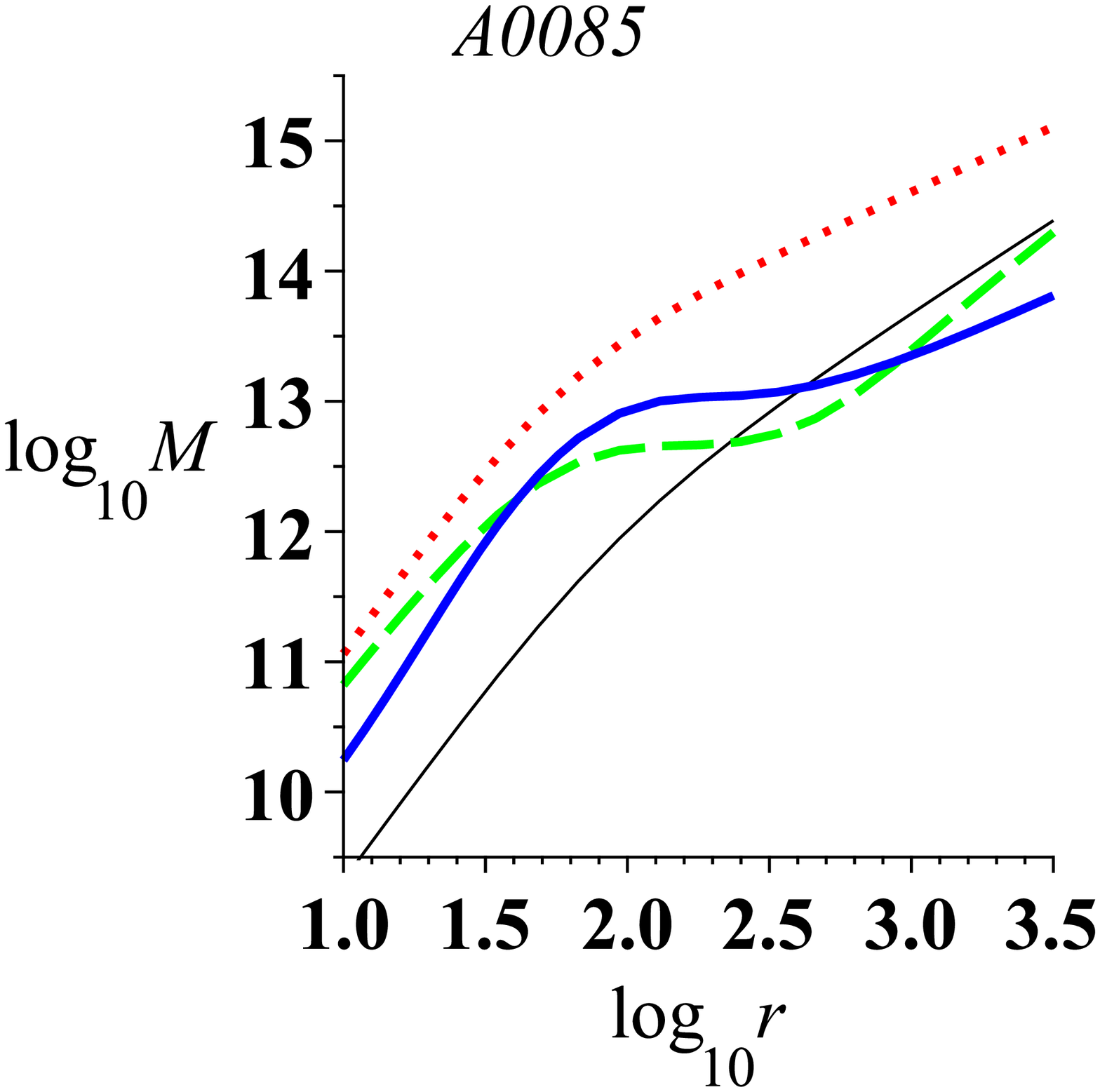}
\hskip 0.01\linewidth
\includegraphics[width=0.24\linewidth]{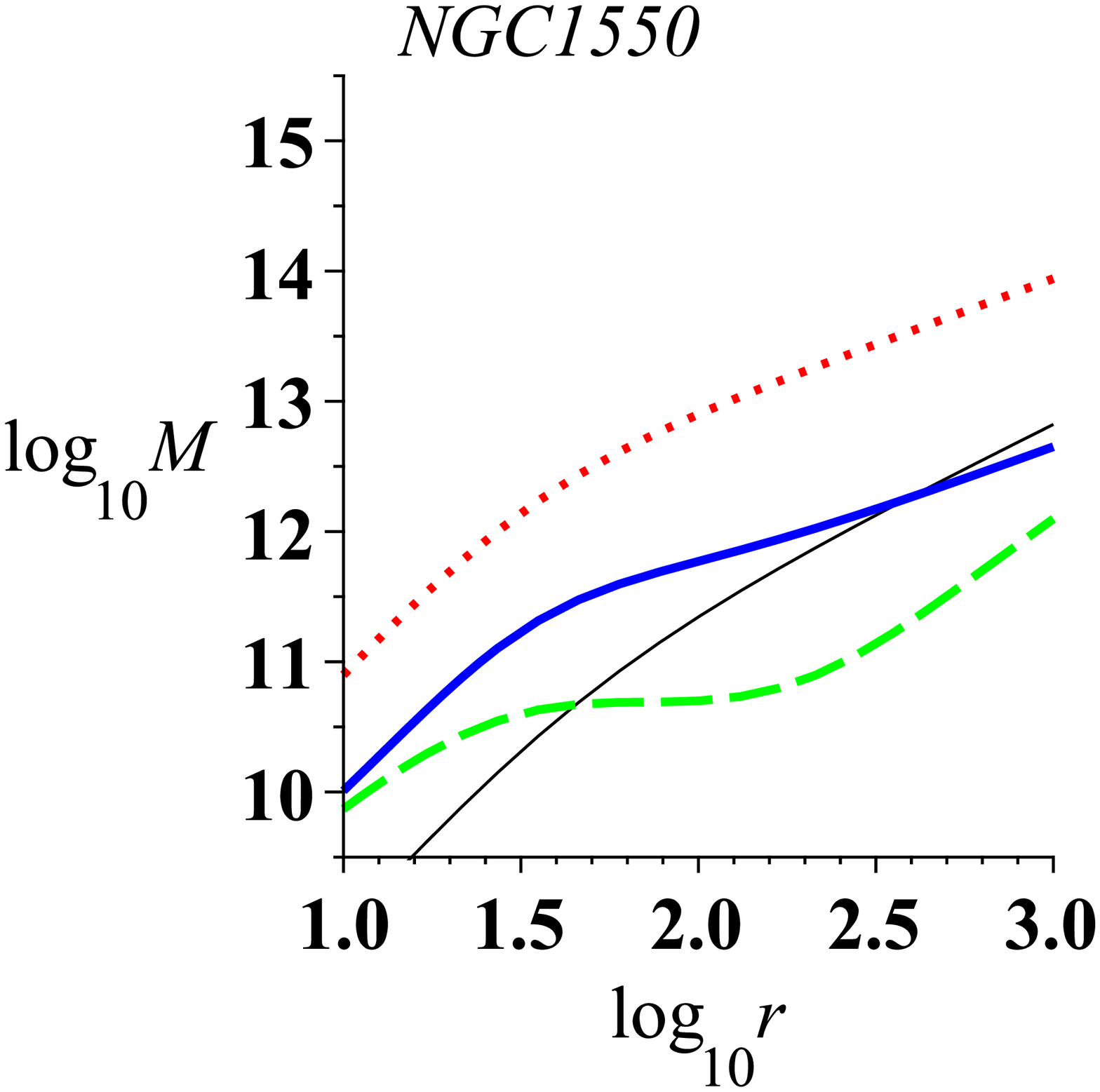}
\hskip 0.01\linewidth
\includegraphics[width=0.24\linewidth]{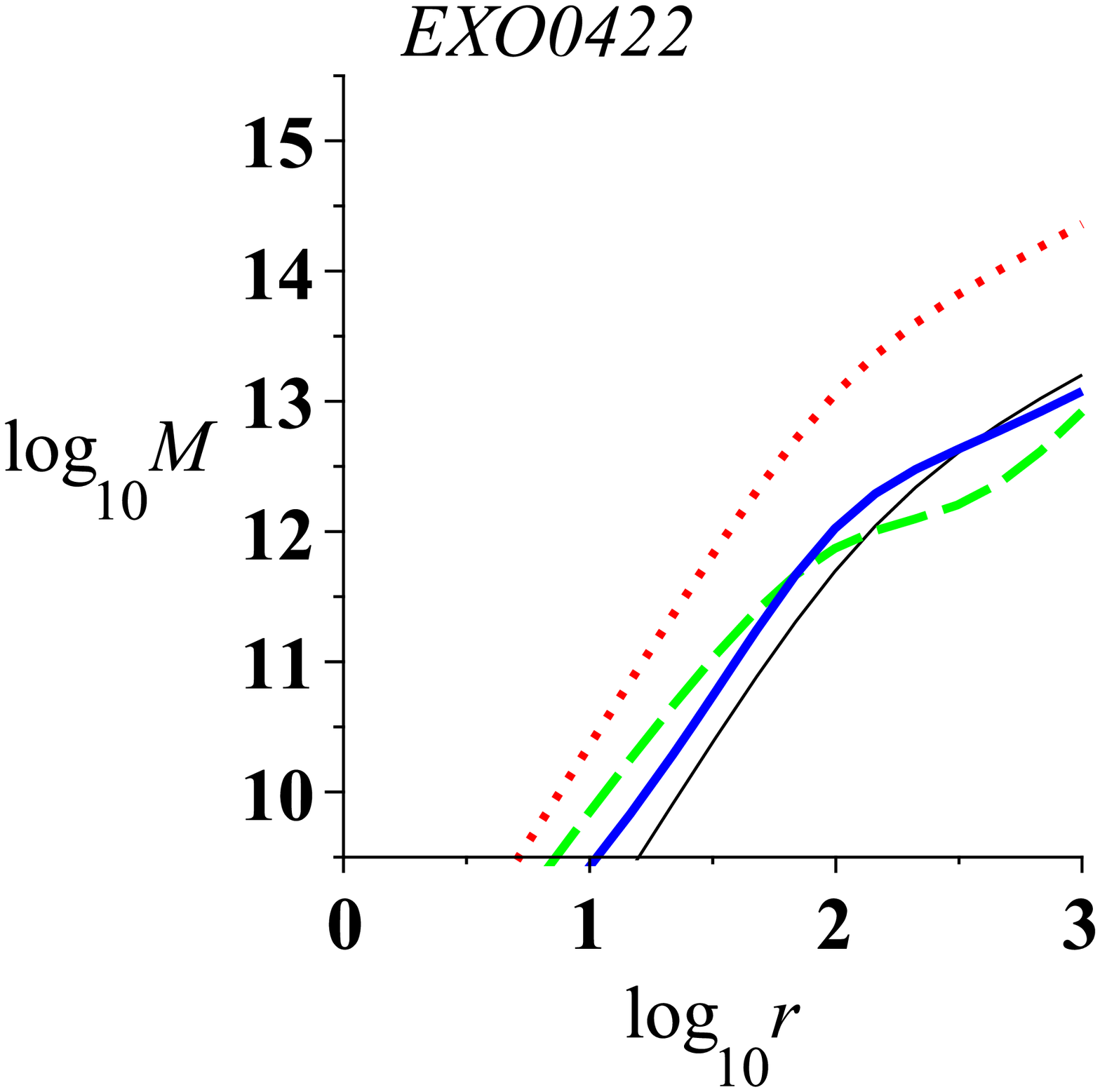}
\hskip 0.01\linewidth
\includegraphics[width=0.24\linewidth]{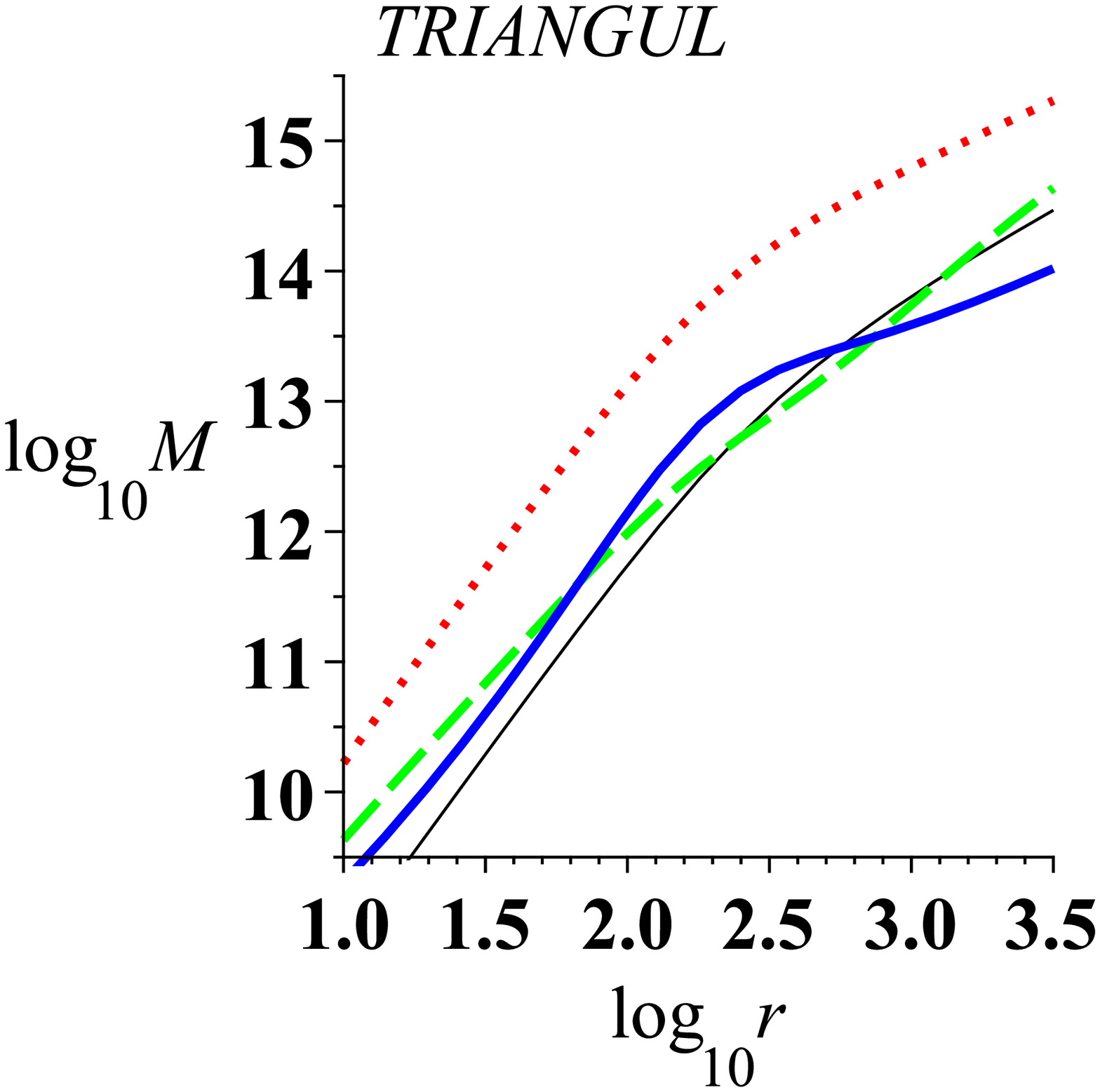}
\caption{A small sample of galaxy clusters studied in \cite{Brownstein2006b}. Thin (black) solid line is the mass profile estimate from \cite{RB2002}. Thick (blue) solid line is the mass profile estimate from the parameter-free solution in \cite{Moffat2007e}. Dashed (green) line is the result published in \cite{Brownstein2006b}, while the dotted (red) line is the Newtonian mass profile estimate. Radial distances are measured in kpc, masses in $M_\odot$.}
\label{fig:clusters}
\end{figure}

MOG has also been used to fit a large sample of X-ray mass profile cluster data (106 clusters) \cite{Moffat2007e,Brownstein2006b}; some representative clusters are shown in Fig~\ref{fig:clusters}.

The merging clusters 1E0657-56, known as the ``Bullet Cluster'' (at a redshift of $z=0.296$) is claimed to prove empirically the existence of dark matter \cite{Clowe2006,Bradac:2006er}. Due to the collision of two clusters, the dissipationless stellar component and the X-ray emitting plasma are spatially segregated. The claim is that the gravitational lensing maps of the cluster show that the gravitational potential does not trace the plasma distribution---the dominant baryonic mass component---but rather approximately traces the distribution of galaxies. MOG explains the spatial offset of the center of the total mass from the center of the baryonic mass peaks without nonbaryonic dark matter \cite{Brownstein2007}.

On smaller scales, MOG predicts no appreciable deviation from Newtonian gravity. In particular, MOG predicts velocity dispersion curves for globular clusters that are identical to the Newtonian predictions, in good agreement with observation \cite{Moffat2007a}.

\section{MOG Cosmology}

To study the cosmological consequences of MOG, we adopt a FLRW background spacetime described by the usual metric
\begin{equation}
ds^2=dt^2-a^2(t)\left(\frac{dr^2}{1-kr^2}+r^2d\Omega^2\right).
\end{equation}
Under the usual assumptions of spatial homogeneity and isotropy, the MOG vector field obeys the constraints $\phi_i=0$ ($i=1,2,3$) and $B_{\mu\nu}=0$. The modified Friedmann equations are \cite{Moffat2007e,Moffat2007c}:

\begin{equation}
H^2+\frac{k}{a^2}=\frac{8\pi G\rho}{3}-\frac{4\pi}{3}\left(\frac{\dot{G}^2}{G^2}+\frac{\dot{\mu}^2}{\mu^2}-\dot{\omega}^2-2\frac{V_G}{G^2}-2\frac{V_\mu}{\mu^2}-2V_\omega\right)+\frac{1}{3}\left(G\omega\mu^2\phi_0^2+2\omega GV_\phi\right)+\frac{\Lambda}{3}+H\frac{\dot{G}}{G},\label{eq:FR1}
\end{equation}
\begin{equation}
\frac{\ddot{a}}{a}=-\frac{4\pi G}{3}(\rho+3p)+\frac{8\pi}{3}\left(\frac{\dot{G}^2}{G^2}+\frac{\dot{\mu}^2}{\mu^2}-\dot{\omega}^2+\frac{V_G}{G^2}+\frac{V_\mu}{\mu^2}+V_\omega\right)+\frac{2}{3}\left(\omega GV_\phi-G\omega\mu^2\phi_0^2\right)+\frac{\Lambda}{3}+H\frac{\dot{G}}{2G}+\frac{\ddot{G}}{2G}-\frac{\dot{G}^2}{G^2}.
\label{eq:FR2}
\end{equation}

\begin{figure}[t!]
\includegraphics[width=0.45\linewidth]{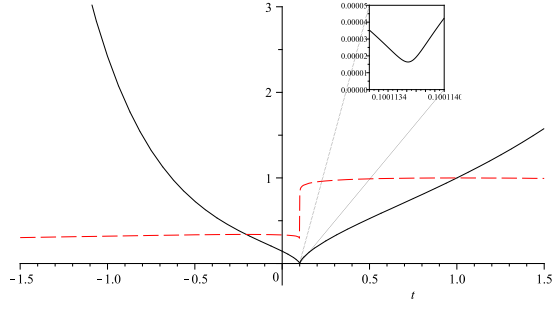}
\caption{The MOG ``bouncing'' cosmology. The horizontal axis represents time, measured in Hubble units of $H_0^{-1}$. The solid (black) line is $a/a_0$, the scale factor normalized to the present epoch. The dashed (red) line is $G/G_0$. The inset shows details of the bounce, demonstrating that a smooth bounce occurs even as the matter density of the universe is more than $10^{14}$ times its present value.}
\label{fig:aG}
\end{figure}

As in the static, spherically symmetric case, it is possible to obtain an exact numerical solution of the MOG field equations for cosmology. Notably, given appropriately chosen initial conditions, MOG is shown to admit a ``bouncing'' cosmology (Fig.~\ref{fig:aG}) \cite{Moffat2007c,Moffat2007e}.

\begin{figure}[t!]
\includegraphics[width=0.45\linewidth]{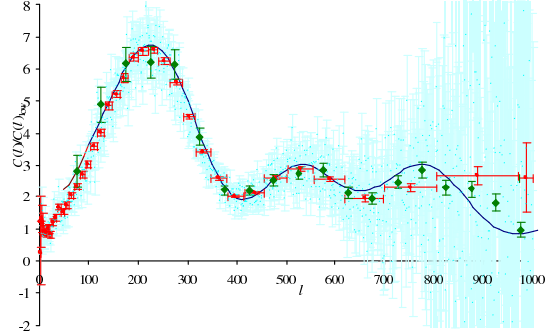}
\caption{The angular CMB power spectrum showing good agreement between the MOG prediction and WMAP-3 and Boomerang data \cite{Moffat2007c}.}
\label{fig:CMB}
\end{figure}

\begin{figure}[t!]
\includegraphics[width=0.4\linewidth]{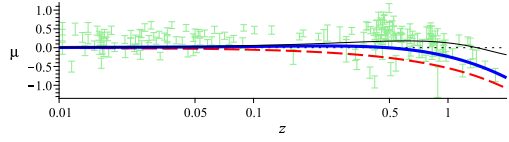}
\caption{the luminosity-distance relationship of type Ia supernovae, with the MOG prediction shown with a thick (blue) line. Thin (black) line is the $\Lambda$CDM prediction; dashed (red) line is a flat Einstein-de Sitter universe, while the horizontal dashed black line corresponds to an empty universe with no deceleration \cite{Moffat2007c}.}
\label{fig:SN}
\end{figure}

\begin{figure}[t!]
\includegraphics[width=0.45\linewidth]{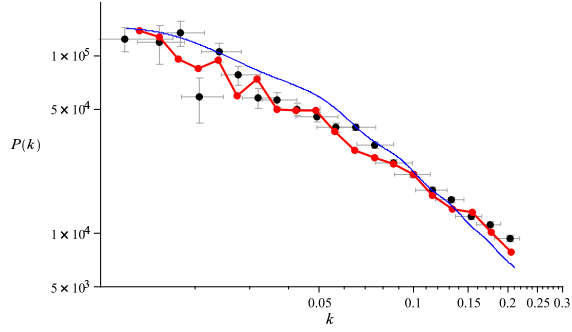}
\caption{MOG (thick red line) shows agreement with the SDSS luminous red galaxy survey mass power spectrum, perhaps even superior to the $\Lambda$CDM prediction (thin blue line) \cite{Moffat2007c}.}
\label{fig:baryon}
\end{figure}

In \cite{Mukhanov2005}, Mukhanov presented an analytic calculation of the cosmic microwave spectrum. Before embarking on the tedious exercise of modifying numerical codes to compute a precision MOG prediction, we considered it prudent to use Mukhanov's simpler approach to check if MOG can reproduce the acoustic peaks in the CMB. We found that indeed this was the case: after the appropriate substitution of model parameters, the resulting acoustic power spectrum shows good agreement with the data (Fig.~\ref{fig:CMB}). This result was obtained assuming a baryon density that is 4\% of the critical density ($\Omega_b=0.04$).

The role played by CDM in the standard $\Lambda$CDM model is replaced in MOG by the significant deepening of the gravitational wells before recombination with $G_\mathrm{eff}\simeq 7G_N$, which traps the baryons. This reduces the baryon dissipation due to the photon coupling pressure (Silk damping) and the third and higher peaks in the acoustical oscillation spectrum are not suppressed due to finite thickness and baryon drag effects. The effective baryon density $\Omega_{b\mathrm{eff}}=(1+\alpha)\Omega_b\simeq 7\Omega_b\simeq 0.3$ dominates before recombination and {\bf we fit the acoustical spectrum without a collisionless dark matter component}.

The deceleration parameter $q$ can be expressed from the MOG Friedmann equations as
\begin{equation}
q=\frac{1}{2}(1+3w_\mathrm{eff})\left(1+\frac{k}{\dot{a}^2}\right)-\frac{1}{2}(1+w_\mathrm{eff})\frac{\Lambda}{H^2}.
\end{equation}
{\bf The effective MOG equation of state $w_\mathrm{eff}$ is between $-1$ and $0$ \cite{Moffat2007c}, and descends below $-1/3$ when the universe is about $2/3$ its present age. This allows $q$ to be negative, resulting in an accelerating universe even when $\Lambda=0$.} Figure~\ref{fig:SN} compares the MOG and $\Lambda$CDM deceleration parameters to Type Ia supernova data.

MOG can also account for the mass power spectrum that is obtained from large-scale galaxy surveys. This power spectrum is a result of gravitational instabilities that lead to the growth of density fluctuations. Schematically, the growth of density fluctuations $\delta$ is governed by the equation
\begin{equation}
\ddot\delta+[\mathrm{Pressure}-\mathrm{Gravity}]\delta=0.
\end{equation}
If the pressure is low, $\delta$ grows exponentially, while if pressure is high, $\delta$ oscillates with time. The effective increase in the gravitational constant $G_\mathrm{eff}\simeq 7G_N$ produces growth at earlier times before recombination, mimicking the effects of cold dark matter. {\bf Thus, the observed power spectrum can be predicted by MOG without cold dark matter.} This is demonstrated in Fig.~\ref{fig:baryon}, which compares the MOG and $\Lambda$CDM predictions against actual data.

\section{MOG Verifiable Predictions}

The preceding sections demonstrated that MOG has the ability to reproduce key results of $\Lambda$CDM cosmology without resorting to the use of the cold dark matter paradigm.

On the other hand, some specific predictions of MOG may be at odds with prevailing cosmological models, leading to possible observational tests that can be used to distinguish them from each other.

The first such prediction concerns the mass power spectrum. As we have seen, MOG reproduces well the observational results obtained from large scale galaxy surveys. However, the MOG prediction differs from the $\Lambda$CDM prediction in an essential way: the mass power spectrum predicted by MOG is characterized by unit oscillations (baryonic oscillations) which are largely absent from the $\Lambda$CDM prediction due to the dominance of cold dark matter.

Unfortunately, the finite size of samples and the associated window functions that are used to bin galaxy survey data are masking any such oscillations in the power spectrum. This is reflected in Fig.~\ref{fig:baryon}. However, these oscillations may become observable as the galaxy surveys increase in size and the window functions narrow in size.

\begin{figure}[t!]
\includegraphics[height=0.5\linewidth,angle=270]{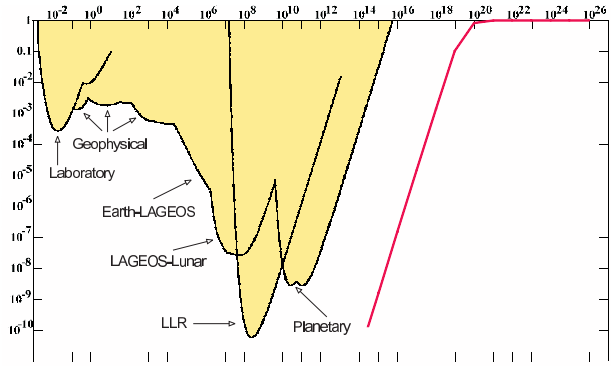}
\caption{Predictions of the Yukawa-parameters from the MOG field equations are not in violation of solar system and laboratory constraints. Predicted values of $\lambda$ (horizontal axis, in m) vs. $|\alpha_Y|$ are indicated by the solid red line. Plot adapted from \cite{Adelberger2003}.}
\label{fig:noses}
\end{figure}

The weak field MOG modification to the Newtonian acceleration law if of a Yukawa type. However, within the solar system, the smallness of the MOG parameters means that the theory predicts only extremely tiny deviations from the Einstein prediction. The sensitivity of observations must improve by several orders of magnitude before these deviations can be observed (see Fig.~\ref{fig:noses}). In the future, however, such sensitive observations may become possible (e.g., Mars laser ranging) and MOG can be tested via solar system experiments.

\begin{figure}[t!]
\includegraphics[width=0.45\linewidth]{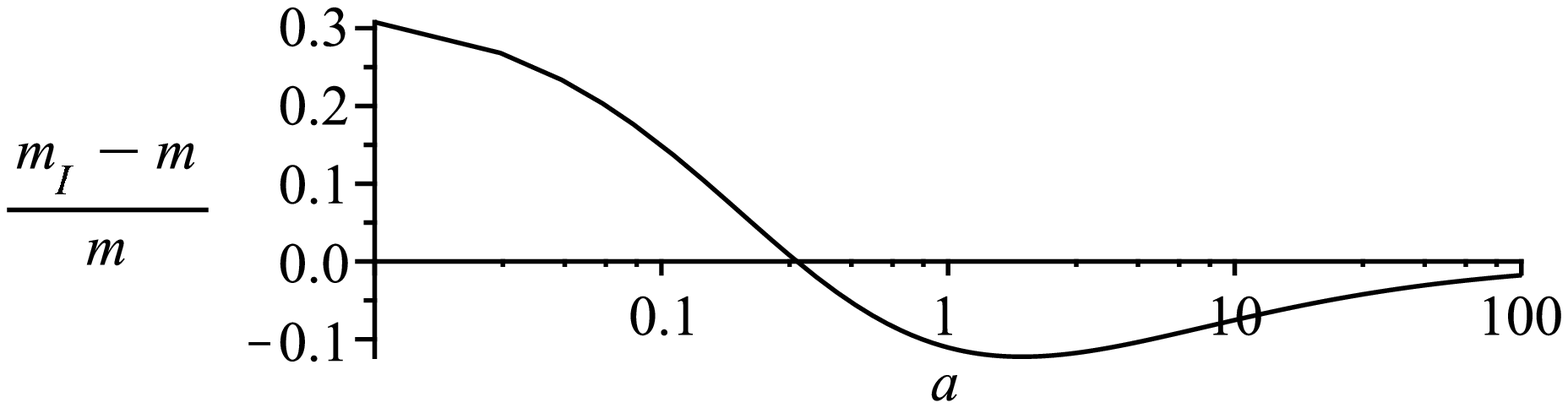}
\hskip 0.1\linewidth
\includegraphics[width=0.45\linewidth]{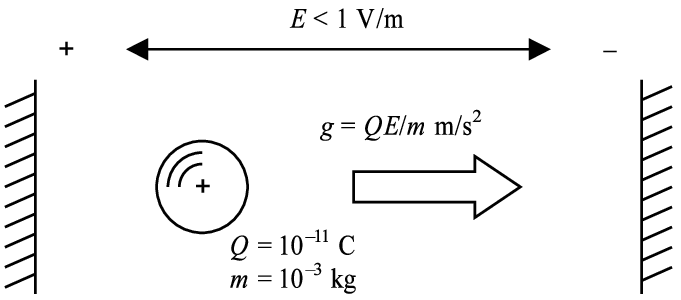}
\caption{Does MOG violate the weak equivalence principle for very small accelerations? The horizontal axis in the plot (left) is the acceleration $g$, measured in units of the cosmic acceleration $cH_0\simeq 7\times10^{-10}$~m/s$^2$. The vertical axis shows the predicted difference between inertial mass $m_I=–F(g)/g$ and passive gravitational mass $m$. The schematic on the right depicts a simple experiment that can be used to verify the validity of the force law $F=mg$ for very small accelerations. With the values presented here, a measurement of a deflection of $\sim 1.8$~mm over the course of ten minutes with an accuracy better than 10\% is required, in order to measure the deficit in inertial mass. A smaller acceleration (corresponding to $E\simeq 0.01$~V/m) could be used to measure an excess in inertial mass of up to ~30\%.}
\label{fig:inertia}
\end{figure}

MOG does not satisfy Birkhoff's theorem. It can realize Mach's principle and explain the origin of inertia \cite{Moffat2007d}. The inertial force arises as the influence of distant matter in the universe. Yet MOG also predicts a small deviation from d'Alembert's law of inertia at very low accelerations in systems that are non-accelerating with respect to the cosmic background (Fig.~\ref{fig:inertia}). This deviation may be verifiable by either existing or planned space-based gravitational experiments.

\section{Conclusions}

A stable and self-consistent modified gravity (MOG) is constructed from a pseudo-Riemannian geometry and a skew field obtained from the curl of a massive vector field (phion field) (STVG). {\bf The field equations are derived from a fully relativistic action principle.} The static spherically symmetric solution of the field equations yields a modified Newtonian acceleration law with a distance scale dependence. The gravitational constant $G$, the effective mass and the coupling strength of the skew field run with distance scale $r$.

A fit to galaxy rotations curves is obtained with only M/L as a free parameter without exotic dark matter. The mass profiles of X-ray galaxy clusters are also successfully fitted for those clusters that are isothermal. Similarly, a fit to the Bullet Cluster 1E0657-56 data can be achieved with the running of the gravitational constant $G$ without nonbaryonic dark matter.

The CMB power spectrum acoustical peaks data including the third peak can be fitted with the density parameter $\Omega_{m\mathrm{eff}}=8\pi G_\mathrm{eff}\rho/3H^2$ without having to resort to the postulate of cold dark matter. The power spectrum for growth of fluctuations and the formation of galaxies and clusters can also be incorporated in MOG without dark matter. Furthermore, MOG offers a possible explanation for the acceleration of the universe without Einstein's cosmological constant, while leaving open the possibility of a ``bouncing'' cosmology.

In addition to these successful explanations of existing observations, MOG also offers definite predictions that may be used to distinguish it from other theories, notably the $\Lambda$CDM ``standard model'' of cosmology, in the foreseeable future.


\begin{theacknowledgments}
The research was partially supported by National Research Council of Canada. Research at the Perimeter Institute for Theoretical Physics is supported by the Government of Canada through NSERC and by the Province of Ontario through the Ministry of Research and Innovation (MRI).
\end{theacknowledgments}



\bibliographystyle{aipproc}   

\bibliography{refs}

\IfFileExists{\jobname.bbl}{}
 {\typeout{}
  \typeout{******************************************}
  \typeout{** Please run "bibtex \jobname" to optain}
  \typeout{** the bibliography and then re-run LaTeX}
  \typeout{** twice to fix the references!}
  \typeout{******************************************}
  \typeout{}
 }

\end{document}